\documentclass[aps, prd, onecolumn, tightenlines, notitlepage, superscriptaddress, nofootinbib, preprintnumbers, floatfix,showkeys,11pt,altaffilletter]{revtex4-2}

\usepackage{upgreek}

\usepackage[official]{eurosym}
\usepackage[normalem]{ulem}
\usepackage{amstext}
\usepackage{graphicx}
\usepackage{url}
\usepackage{color}
\usepackage{ulem}
\usepackage[version=4]{mhchem}
\usepackage[utf8]{inputenc}
\usepackage{fontawesome}
\usepackage{yfonts}
\usepackage{slashed}

\pdfoutput=1
\usepackage{textcomp}
\usepackage{comment}
\usepackage{yfonts}
\usepackage{epsfig,amsfonts,mathrsfs,graphicx,color,slashed,multirow}
\usepackage{latexsym,graphicx,slashed,color,enumerate,url,cancel,gensymb}
\usepackage{textcomp}

\usepackage[x11names]{xcolor}
\usepackage[colorlinks,pdfstartview=FitV,breaklinks=true]{hyperref}
\usepackage{booktabs}
\usepackage{adjustbox}
\usepackage{textgreek} 
\usepackage{caption, subcaption} 
\usepackage{booktabs} 
\usepackage{float}

\usepackage{lmodern}
\usepackage{ae,aecompl}
\usepackage{appendix}
\usepackage{orcidlink}
\usepackage{nccmath} 
\usepackage{textcomp}
\allowdisplaybreaks 


\definecolor{byzantium}{rgb}{0.44, 0.16, 0.39}

\makeatletter
    \newcommand{\colorboxed}[3][white]{\fcolorbox{#2}{#1}{\m@th$\displaystyle#3$}}
\makeatother

\AtBeginDocument{\hypersetup{citecolor=byzantium,linkcolor=byzantium,urlcolor=byzantium}}
\usepackage{appendix}

\begin{document}

\title{{\LARGE Neutrino magnetic moments: effective versus fundamental parameters}}

\author{Christoph A. Ternes~\orcidlink{0000-0002-7190-1581}}
\email{christoph.ternes@lngs.infn.it}
\affiliation{Istituto Nazionale di Fisica Nucleare (INFN), Laboratori Nazionali del Gran Sasso, 67100 Assergi, L’Aquila (AQ), Italy
}
\author{Mariam T\'ortola~\orcidlink{0000-0002-5855-2671}}\email{mariam@ific.uv.es}
\affiliation{Instituto de F\'{i}sica Corpuscular (CSIC-Universitat de Val\`{e}ncia), Parc Cient\'ific UV C/ Catedr\'atico Jos\'e Beltr\'an, 2 E-46980 Paterna (Valencia), Spain}
\affiliation{Departament de F\'isica  Te\'orica,  Universitat  de  Val\`{e}ncia, Spain}


\begin{abstract}

The search for neutrino magnetic moments offers a valuable window into physics beyond the Standard Model.
However, a common misconception arises in the interpretation of experimental results: the assumption that the so-called effective neutrino magnetic moment is a universal, experiment-independent quantity. In reality, this effective parameter depends on the specific characteristics of each experiment, including the neutrino source, flavor composition or energy spectrum. As a result, the effective magnetic moment derived from solar neutrino data differs fundamentally from that obtained in reactor or accelerator-based experiments. Treating these quantities as directly comparable can lead to misleading conclusions. 
In this work, we clarify the proper definition of the effective neutrino magnetic moment in various experimental contexts and discuss the implications of this misconception for global analyses and theoretical interpretations.

\end{abstract}

\maketitle


\section{Introduction}

In recent years, neutrino electromagnetic properties~\cite{Schechter:1981hw, Nieves:1981zt, Shrock:1982sc, Kayser:1982br} have become a subject of intense research. 
Both experimental collaborations~\cite{Ahrens:1990fp,Allen:1992qe,LSND:2001akn,DONUT:2001zvi,MUNU:2005xnz,TEXONO:2006xds,Beda:2012zz,Borexino:2017fbd,PandaX-II:2020udv,XENON:2022ltv} and phenomenologists~\cite{AtzoriCorona:2022jeb,A:2022acy,Khan:2022bel,AtzoriCorona:2022qrf,Coloma:2022avw,Liao:2022hno,Akhmedov:2022txm,Khan:2022akj,DeRomeri:2022twg,Coloma:2022umy,Giunti:2023yha,Frigerio:2024jlh,DeRomeri:2024hvc,DeRomeri:2025csu,AtzoriCorona:2025ygn} have analyzed available data in search of signs of such properties. 
Particular attention has been given to neutrino magnetic and electric transition and dipole moments,  which can arise in many extensions of the Standard Model (SM) that account for neutrino masses~\cite{Giunti:2014ixa}.
The sensitivities achieved in current experiments remain several orders of magnitude above the values predicted by the minimal extension of the SM with right-handed Dirac neutrinos~\cite{Fujikawa:1980yx,Pal:1981rm,Shrock:1982sc}. Nevertheless, in more sophisticated theoretical frameworks -see, for example, Ref.~\cite{Babu:2020ivd}- the predicted neutrino magnetic moments can be significantly larger and may be within reach of current or upcoming experimental efforts. For a comprehensive overview, see the review in \cite{Giunti:2014ixa}.

Neutrino electromagnetic properties are typically probed by searching for deviations in the low-energy  spectra of experiments that measure nuclear or electron recoils via coherent elastic neutrino-nucleus scattering (CE$\nu$NS) or elastic neutrino-electron scattering (E$\nu$ES), respectively. Due to the helicity flipping nature of the interaction, the magnetic moment contribution adds incoherently to the SM cross sections, leading to a characteristic enhancement at low energies. 
The magnetic moment cross section is given by~\cite{Vogel:1989iv}
\begin{equation}
  \label{eq:xsec_magmom}
  \left. \frac{d\sigma}{dT}\right|^\mathrm{MM}
\propto
\dfrac{ \pi \alpha^2_\mathrm{EM} }{ m_{e}^2 }
\left( \dfrac{1}{T} - \dfrac{1}{E_\nu} \right)
\left| \dfrac{\mu_{\nu_{\ell}}}{\mu_{\text{B}}} \right|^2 \, ,
\end{equation}
where $\alpha_\mathrm{EM}$ is the fine-structure constant, $m_e$ the electron mass, $T$ the nuclear or electron recoil energy, $E_\nu$ the neutrino energy, $\mu_{\nu_{\ell}}$ is the effective magnetic moment for $\nu_{\ell}$, and $\mu_{\text{B}} = e/(2m_e)$ is the Bohr magneton. In the case of CE$\nu$NS experiments, this cross section must be multiplied by the nuclear form factor squared and the proton number squared, whereas in E$\nu$ES experiments, it is scaled by the effective number of ionizable electrons.

Experimental data are commonly used to set limits on the effective neutrino magnetic moment, $\mu_{\nu_{\ell}}$. These bounds are often compared across different experiments and neutrino flavors, without recognizing that $\mu_{\nu_{\ell}}$ is not a fundamental quantity. Instead, it is an effective parameter that depends on several factors, including the experimental setup, the neutrino energy spectrum, the neutrino flavor, and even whether neutrinos are Dirac or Majorana particles.
This issue has been previously discussed in the literature (see, e.g., Refs.~\cite{Grimus:2002vb,Canas:2015yoa} and the recent Ref.~\cite{AristizabalSierra:2021fuc}), yet some confusion persists in the field. 
The aim of this article is to clarify the connections among the different effective magnetic moments $\mu_{\nu_{\ell}}$ and to update the bounds on the underlying physical parameters, which are independent of experimental assumptions. 
Based on the most stringent limits on these fundamental parameters, we define a benchmark "playground" that future experiments must reach in order to surpass current constraints.

The paper is structured as follows: In Section \ref{sec:theoretical-framework} we introduce our notation and explain how the effective magnetic moments are related to the underlying fundamental parameters. Section~\ref{sec:res} reviews the constraints obtained from different experimental setups. In Section~\ref{sec:implications} we use the most stringent bounds on the fundamental parameters to derive limits on the individual effective magnetic moments. Finally, Section~\ref{sec:conc} presents our conclusions.

\section{Theoretical framework}
\label{sec:theoretical-framework}

The effective Hamiltonian that accounts for electromagnetic interactions depends on the Majorana or Dirac nature of neutrinos. In this paper, we will mainly be interested in Majorana neutrinos. In this case, the Hamiltonian can be written as~\cite{Schechter:1981hw,Schechter:1981cv}
\begin{equation}
H_{\text{EM}}^{\mathrm{M}} = -\frac{1}{4} \nu_L^\text{T} C^{-1} \lambda^{\mathrm{M}} \sigma^{\alpha \beta} \nu_L F_{\alpha \beta} + \mathrm{h.c.} \, ,
\label{Hamiltonian:Majorana}
\end{equation}
where $F_{\alpha \beta}$ is the electromagnetic field tensor,  $\lambda^{\mathrm{M}} = \mu^{\mathrm{M}} - i d^{\mathrm{M}}$ is an antisymmetric complex matrix, with $\mu^{\mathrm{M}}$ and $d^{\mathrm{M}}$ being the magnetic and electric moments for Majorana neutrinos. Due to its antisymmetric nature, $\lambda^{\mathrm{M}}$ is characterized by three complex (or six real) parameters.
The corresponding Hamiltonian for the case of Dirac neutrinos is instead given by
\begin{equation}
H_{\text{EM}}^{\mathrm{D}} = \frac{1}{2} \bar{\nu}_R \lambda^{\mathrm{D}} \sigma^{\alpha \beta} \nu_L F_{\alpha \beta} + \mathrm{h.c.}
\label{Hamiltonian:Dirac}
\end{equation}
This time, $\lambda^{\mathrm{D}} = \mu^{\mathrm{D}} - i d^{\mathrm{D}}$  is an arbitrary complex matrix, restricted only to the hermiticity of the Hamiltonian, implying $\mu^{\mathrm{D}} = (\mu^{\mathrm{D}})^\dagger$ and $d^{\mathrm{D}} = (d^{\mathrm{D}})^\dagger$.
Note that, due to the different structure of the Hamiltonian, neutrino electromagnetic properties can potentially be used to distinguish between the Dirac and Majorana neutrino nature. 
A key difference lies in the antisymmetric nature of $\lambda^{\mathrm{M}}$ for Majorana neutrinos, which enforces vanishing diagonal elements: $\mu^\mathrm{M}_{ii} = d^\mathrm{M}_{ii} = 0$. In contrast, for Dirac neutrinos, diagonal magnetic and electric dipole moments are allowed. 
Numerical estimates for the magnetic and electric moments have been obtained within the minimal $\mathrm{SU(2)_L \otimes U(1)_Y}$ model~\cite{Shrock:1982sc}. In this scenario, however, the predicted values are extremely small, well below the reach of current and near-future neutrino experiments. In contrast, in extended models beyond the SM, significantly larger moments can arise, with values potentially within the sensitivity range of present or upcoming experiments~\cite{Giunti:2014ixa,Babu:2020ivd}.
In this work, we remain agnostic about the specific underlying model that generates these moments and focus on the study of  Majorana transition magnetic moments (TMM), denoted by $\mu_{ij}^{\mathrm{M}}$. For simplicity, we will omit the superscript "M" referring to Majorana neutrinos from here on.

The effective neutrino magnetic moment observed in a given experiment depends on the underlying neutrino magnetic moment matrix and the amplitudes of the positive and negative helicity neutrino states, denoted by the three-component vectors $a_{+}$ and $a_{-}$, respectively.  
In the flavor basis, one finds~\cite{Grimus:2000tq} 
\begin{equation}
\left(\mu_\nu^{F} \right)^2 = a_{-}^\dagger \lambda^\dagger \lambda a_{-} + a_{+}^\dagger \lambda \lambda^\dagger a_{+} \, ,
\label{eq:TMM-flavor}
\end{equation}
where $\lambda$ is the TMM matrix in the flavor basis. In the mass basis, the TMM matrix, $\tilde{\lambda}$, as well as the corresponding three-component neutrino vector states, $\tilde{a}_{-}$ and $\tilde{a}_{+}$, can be obtained from the following transformations
\begin{equation}
\tilde{a}_{-} = U^\dagger a_{-}, \qquad  \tilde{a}_{+} = U^\text{T} a_{+}, \qquad \tilde{\lambda} = U^\text{T} \lambda U \, , 
\end{equation}
where $U$ is the leptonic mixing matrix. With these elements one can define  the effective neutrino magnetic moment in the mass basis as~\cite{Grimus:2002vb}, 
\begin{equation}
\left(\mu_\nu^{M} \right)^2 = \tilde{a}_{-}^\dagger \tilde{\lambda}^\dagger \tilde{\lambda} \tilde{a}_{-} + \tilde{a}_{+}^\dagger \tilde{\lambda} \tilde{\lambda}^\dagger \tilde{a}_{+} \, ,
\label{eq:TMM-mass}
\end{equation} 
As mentioned above, three complex quantities are required to parametrize the TMM, so we can write the TMM matrices in the flavour and mass basis as
\begin{equation}
\lambda = \left( \begin{array}{ccc}
0 & \Lambda_\tau & - \Lambda_\mu \\
- \Lambda_\tau &  0 & \Lambda_e \\
\Lambda_\mu & - \Lambda_e & 0
\end{array} \right), \qquad
\tilde{\lambda} = \left( \begin{array}{ccc}
0 & \Lambda_3 & - \Lambda_2 \\
- \Lambda_3 &  0 & \Lambda_1 \\
\Lambda_2 & - \Lambda_1 & 0
\end{array} \right) \, ,
\label{NMM:matrix}
\end{equation}
where we have introduced the complex parameters~\cite{Grimus:2002vb}
\begin{equation}
\Lambda_{\alpha} = |\Lambda_\alpha| e^{i \varphi_\alpha}, \qquad \Lambda_{i} = |\Lambda_i| e^{i \varphi_i}\, .
\label{eq:def-lambda}
\end{equation}
As argued in Ref.~\cite{Grimus:2000tq}, among the six CP-violating phases that appear in the neutrino mixing matrix and the TMM matrix, only three are physically relevant. Following the convention adopted in Refs.~\cite{Canas:2015yoa, AristizabalSierra:2021fuc}, together with the $\delta_\mathrm{CP}$ phase in the neutrino mixing matrix, we  consider only two of the phases $\varphi_i$  to be non-zero and assume $\Lambda_2$ to be a real parameter. 

Note that, although we have presented two equivalent expressions for the effective magnetic moment -Eqs.~\eqref{eq:TMM-flavor} and~\eqref{eq:TMM-mass}- in the flavour and mass bases, respectively, the quantity itself is, of course, basis-independent. However, it is evident from those expressions that the specific value  of the effective magnetic moment  depends on the characteristics of each experimental setup. In the next section, when discussing the bounds from different experiments, we will highlight some of these differences.

\section{Experimental bounds on neutrino magnetic moments}
\label{sec:res}

In this section, we review the most relevant bounds on neutrino magnetic moments reported in the literature.
In all cases, we translate the experimental limits on the effective parameter into constraints on the fundamental quantities that enter the Hamiltonian. 
Our focus is on the most stringent bounds obtained from short-baseline (SBL) reactor and accelerator experiments, solar neutrino observations, and dark matter direct detection (DMDD) experiments.
Additional, complementary constraints have been derived 
from CE$\nu$NS measurements in Refs.~\cite{AtzoriCorona:2022qrf,Coloma:2022avw,Liao:2022hno,Khan:2022akj,DeRomeri:2022twg,DeRomeri:2024hvc,DeRomeri:2025csu,AtzoriCorona:2025ygn}. Note, however, that the resulting limits are currently at least an order of magnitude weaker than those discussed here.
On the other hand, astrophysical observations~\cite{Raffelt:1999gv,Viaux:2013lha} provide some of the most stringent constraints  on the neutrino magnetic moment. These  arise from enhanced energy-loss processes in stellar environments, where a non-zero magnetic moment would increase neutrino emission rates (see, e.g.,~\cite{Capozzi:2020cbu}). 
Such bounds are often not emphasized in phenomenological studies focused on terrestrial experiments, as they rely on specific astrophysical models and assumptions that may introduce additional uncertainties. The most stringent bound comes from plasmon decay, for which we have $\mu_{\nu,\text{plasmon}} = \sqrt{\sum_{ij}|\Lambda_{ij}|^2}< 1.2\times10^{-12}~\mu_\text{B}$ at 95\% C.L.
Nonetheless, with DMDD experiments now reaching comparable sensitivities, the omission of astrophysical limits becomes less critical in this context.
A comprehensive list of current bounds on effective magnetic moments can be found in Ref.~\cite{Giunti:2024gec}.

In Section~\ref{sec:SBL_bounds}, we present the leading bounds obtained from SBL  neutrino experiments, specifically GEMMA and LSND. Section~\ref{sec:DMDD_bounds} focuses on constraints from  DMDD experiments, which are sensitive to solar neutrinos: these are compared with the bounds from Borexino. 
For GEMMA, LSND and Borexino we reinterpret the limits on the effective magnetic moments previously obtained in terms of the fundamental parameters entering the Hamiltonian. In contrast, for DMDD experiments we perform a direct fit to the experimental data.
In all analyses below, we minimize over the standard neutrino oscillation parameters incorporating their uncertainties as reported in  Ref.~\cite{deSalas:2020pgw}. Additionally, the new CP-violating phases $\varphi_1$ and $\varphi_3$, as well as the (not displayed) $\Lambda$ parameter, are treated as free parameters and marginalized over in the fits.

\subsection{Bounds from short-baseline experiments}
\label{sec:SBL_bounds}

We begin our discussion with the results  from short-baseline experiments, focusing on GEMMA and LSND. 
GEMMA~\cite{Beda:2012zz}  was a reactor-based neutrino experiment designed to search for neutrino magnetic moments. It operated at the Kalinin Nuclear Power Plant in Russia, approximately 13.9~m from the core of a 3~GW pressurized water reactor. GEMMA employed  a high-purity germanium detector with an active mass of about 1.5~kg,   optimized to detect low-energy electron recoils resulting from E$\nu$ES. 
On the other hand, the Liquid Scintillator Neutrino Experiment (LSND)~\cite{LSND:2001akn} was conducted at Los Alamos National Laboratory between 1993 and 1998, primarily aimed at searching for short-baseline neutrino oscillations. The experiment used an 800~MeV proton beam to produce pions, which decayed to $\mu^+$ and $\nu_\mu$, followed by  $\mu^+$ decays into $e^+$, $\nu_e$ and $\overline{\nu}_\mu$. As a result,  the neutrino beam in LSND consisted of $\nu_\mu$, $\overline{\nu}_\mu$ and $\nu_e$. The detector was sensitive to both E$\nu$ES and inverse beta decay events. 
At 90\% confidence level (CL) the respective collaborations reported the following bounds on the effective neutrino magnetic moment
\begin{eqnarray}
    \mu_\nu &<& 2.9\times10^{-11}\mu_{\text{B}} \quad \text{(GEMMA)}\,, \\
    \mu_\nu &<& 5.7\times10^{-10}\mu_{\text{B}} \quad \text{(LSND)}\label{eq:bound_LSND}\,,
\end{eqnarray}
where the LSND result is obtained from the general bound reported in Ref.~\cite{LSND:2001akn} (see discussion below). In the flavor basis, the effective Majorana TMM relevant for a SBL reactor experiment, with the three-component neutrino and antineutrino vector states
$a_{-}^T = (0,0,0)$ and $a_{+}^T = (1,0,0)$, is given by~\cite{Grimus:2002vb} 
\begin{equation}
\left(\mu_{\nu, \, \text{reactor}}^F \right)^2 = |\Lambda_{\mu}|^2 + |\Lambda_{\tau}|^2 \, ,
\label{TMM-reactor-flavor}
\end{equation}
where $|\Lambda_{\mu}|$ and $|\Lambda_{\tau}|$ are the elements of the neutrino TMM matrix $\lambda$ in Eq.~\eqref{NMM:matrix}.
For LSND, the flavour composition of the neutrino flux implies $a_{-}^T = (1,1,0)$ and $a_{+}^T = (0,1,0)$. The resulting effective magnetic moment in the flavor basis is~\cite{Grimus:2002vb}  
\begin{equation}
\left({\mu_{\nu, \, \text{acceler}}^F} \right)^2 = |{\bf\Lambda}|^2 + |\Lambda_{e}|^2 + 2|\Lambda_{\tau}|^2 - 2|\Lambda_{e}||\Lambda_{\mu}|\cos(\varphi_e-\varphi_\mu) \, ,
\label{TMM-acceler-flavor}
\end{equation}
where $|{\bf\Lambda}|^2 = |\Lambda_{e}|^2 + |\Lambda_{\mu}|^2 + |\Lambda_{\tau}|^2 = |\Lambda_{1}|^2 + |\Lambda_{2}|^2 + |\Lambda_{3}|^2$. 
In the mass basis, these expressions become significantly more complex due to their dependence on the neutrino mixing parameters. For completeness, we refer the reader to Ref.~\cite{Canas:2015yoa}, where the full expressions are provided.
As evident from Eqs.~\eqref{TMM-reactor-flavor} and \eqref{TMM-acceler-flavor}, the expressions for the effective magnetic moments differ in their dependence on the fundamental parameters. They should therefore not be directly compared.

\begin{figure}[!t]
\centering
\includegraphics[width=0.49\textwidth]{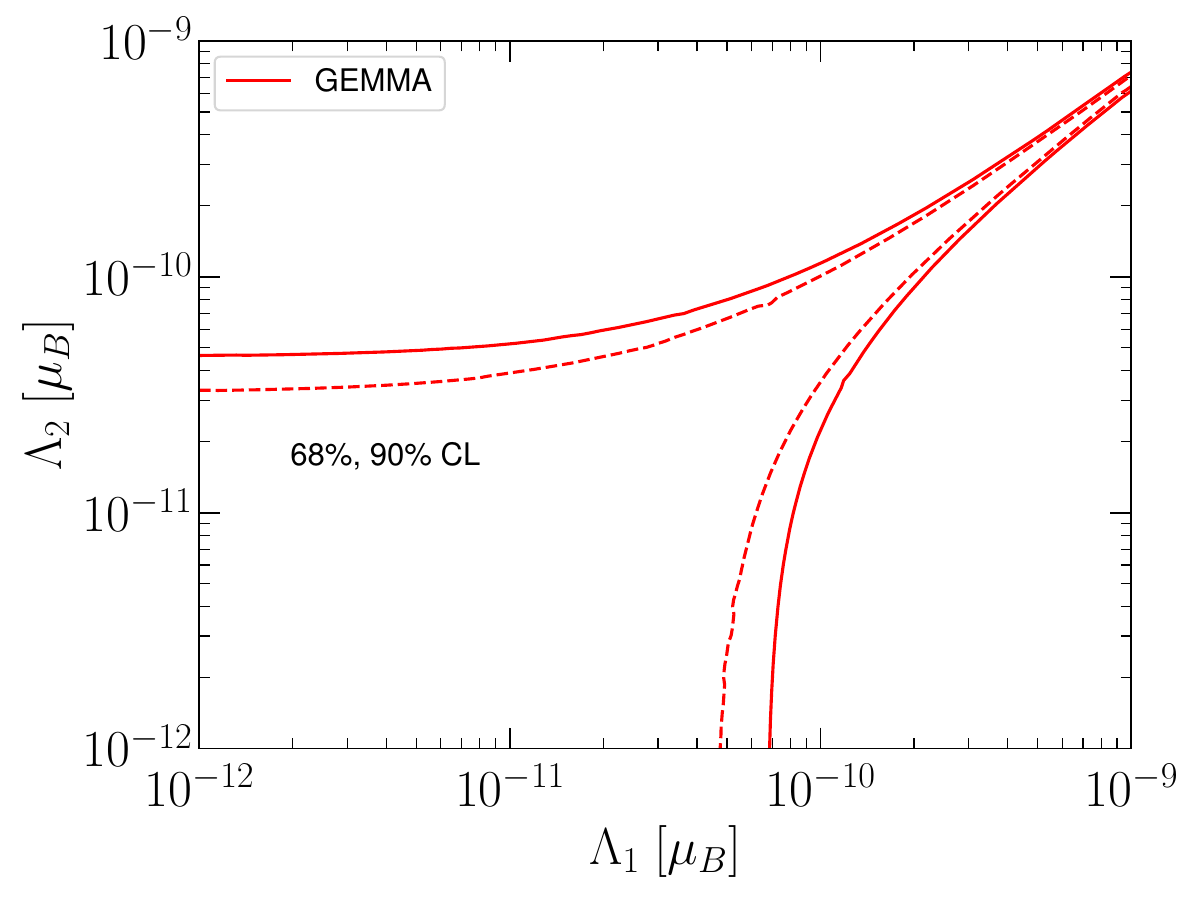}
\includegraphics[width=0.49\textwidth]{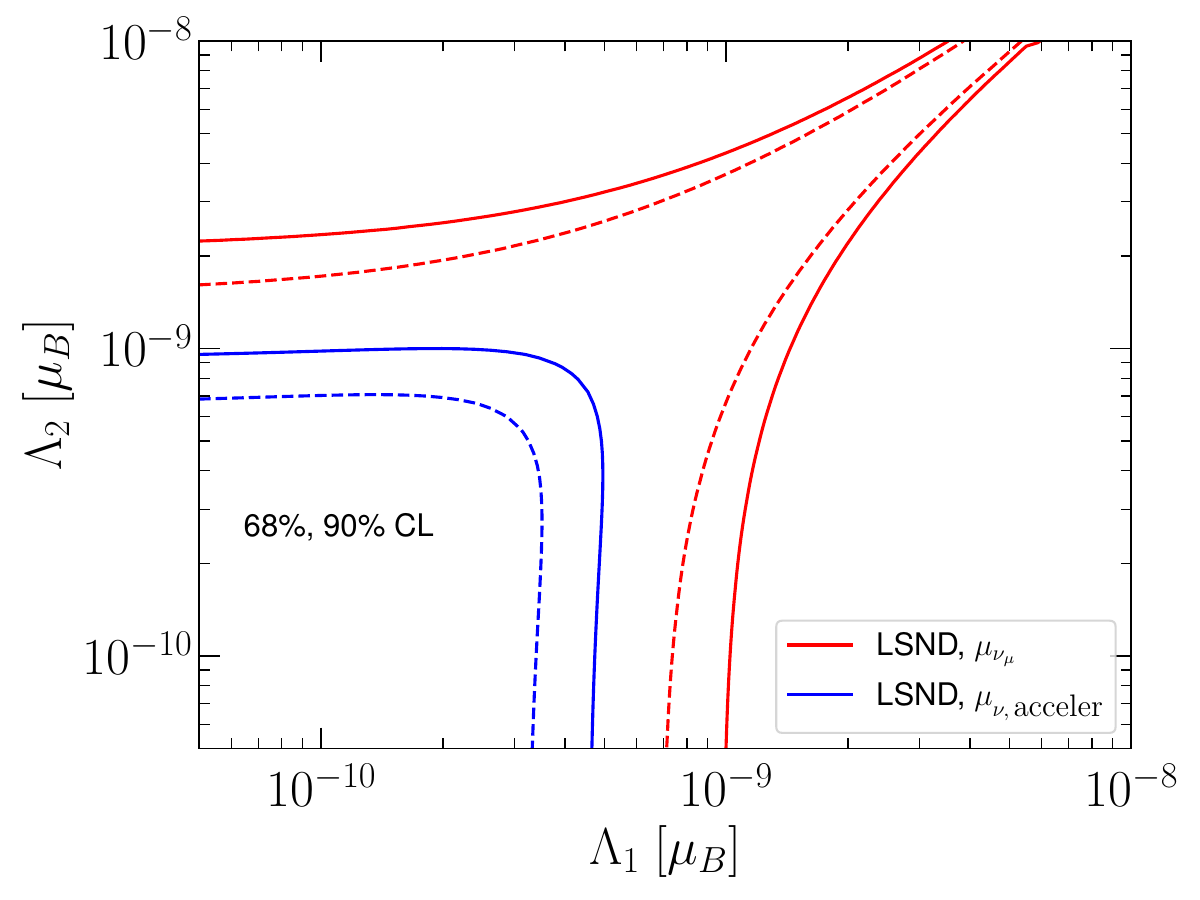}
\caption{The 1$\sigma$ (dashed) and 90\% CL (solid lines) contours in the $\Lambda_1-\Lambda_2$ obtained from GEMMA (left) and LSND (right) data. The LSND region is represented for two different assumptions for the effective magnetic moment. The parameter cancellations for LSND are only present if one attributes the effective magnetic moment only to $\nu_\mu$.}
\label{fig:sbl}
\end{figure}

In the left panel of Fig.~\ref{fig:sbl} we present the results from GEMMA~\cite{Beda:2012zz} in the $\Lambda_1-\Lambda_2$ plane\footnote{Similar contours can be obtained for other combinations of $\Lambda$ parameters in both experiments} showing the 1$\sigma$ and 90\% CL regions for two degrees of freedom. As can be seen, cancellations among the fundamental parameters are possible, meaning that even sizeable transition magnetic moments $\Lambda_i$ can result in small effective magnetic moments in the case of reactor neutrinos. These cancellations occur when the CP-violating phases satisfy $\varphi_1 \approx 0$ and $\varphi_3 \approx \pi$, a feature previously identified and discussed in  Ref.~\cite{AristizabalSierra:2021fuc}. 
In that study,  the authors also 
identified parameter cancellations for LSND. However, it arises only under the assumption that the entire effective magnetic moment originates from interactions involving muon neutrinos\footnote{In Ref.~\cite{LSND:2001akn}, LSND reported a bound on a combination of electron and muon magnetic moments. When this quantity is translated into $\mu_{\nu, \, \text{acceler}}$, we obtain the bound reported in Eq.~\eqref{eq:bound_LSND}. By setting the component due to electron neutrino interactions to zero, one obtains the value commonly quoted in the literature  $\mu_{\nu_\mu}<6.8\times10^{-10}$ at 90\% CL. It should be noted, however, that this is not 
a physical assumption, as there is no combination of fundamental parameters that could lead to such a scenario. Since LSND does not distinguish neutrino flavors, the actual quantity measured is $\mu_{\nu, \, \text{acceler}}$.}. In that case, we have $a_{-}^T = (0,1,0)$ and $a_{+}^T = (0,0,0)$ and the effective magnetic moment is given by
\begin{equation}
\left({\mu_{\nu_\mu}^F} \right)^2 = 
 |\Lambda_{e}|^2 + |\Lambda_{\tau}|^2 \,.
\label{TMM-acc-numu-flavor}
\end{equation}
In this case, achieving the cancellation requires the CP-violating phases to be $\varphi_1 = \pi$ and $\varphi_3 = 0$, opposite to the conditions for reactor neutrinos. As a result, a combined analysis of GEMMA and LSND data cannot simultaneously  lead to cancellations in both experiments. In any case, once the full composition of the LSND neutrino beam—comprising $\nu_\mu$, $\overline{\nu}_\mu$, and $\nu_e$—is properly taken into account, the degeneracy is lifted. This effect is illustrated in the right panel of Fig.~\ref{fig:sbl}, where we compare the confidence contours obtained assuming a pure $\nu_\mu$ beam (red lines) with those from the realistic mixed beam (blue lines).

\subsection{Bounds from solar neutrinos}
\label{sec:DMDD_bounds}

For solar neutrinos,  effective magnetic moments can be probed through elastic scattering  with electrons in detectors such as Super-Kamiokande, Borexino, and also in dark matter direct detection experiments.
The analysis of the recoil electron energy spectrum from solar neutrino scattering provides a powerful tool to constrain the neutrino  magnetic moment. Such analyses have been carried out using data from Super-Kamiokande~\cite{Beacom:1999wx,Grimus:2002vb} and Borexino~\cite{Borexino:2017fbd,Canas:2015yoa,Coloma:2022umy}. In particular, Borexino achieved an order-of-magnitude improvement over previous solar constraints thanks to its sensitivity to lower-energy neutrinos, where the impact of TMMs is more pronounced.
When deriving the effective magnetic moment in the mass basis in this context,
\begin{equation}
\label{eq:nmm_sun0}
(\mu^{M}_{\rm{sol}})^{2}  = \sum_{k,j}\left(\tilde{a}_-^j\right)^\dagger \left(\tilde{\lambda}^\dagger \tilde{\lambda}\right)_{jk}\left(\tilde{a}_-^k\right) = 
\sum_j |\mathbf{\Lambda}|^2 \left(\tilde{a}_-^j\right)^*\left(\tilde{a}_-^j\right) - 
\sum_{j,k} \Lambda_k^* \Lambda_j \left(\tilde{a}_-^j\right)^*\left(\tilde{a}_-^k\right) \,,
\end{equation}
it is essential to account for the fact that neutrinos reach Earth as an incoherent mixture of mass eigenstates due to flavor oscillations during their propagation from the Sun. Therefore, there are no interference terms in the final sum of Eq.~\eqref{eq:nmm_sun0}, and the expression can be rewritten as follows
\begin{equation}
\label{eq:nmm_sun02}
(\mu^{M}_{\rm{sol}})^{2}  =  
|\mathbf{\Lambda}|^2 -\sum_j P^{3\nu}_{ej} |\Lambda_j|^2 \,,
\end{equation}
where we have identified the probability that an electron neutrino produced at the Sun arrives to the Earth as the neutrino mass eigenstate $\nu_j$ by $P_{ej} =\left(\tilde{a}_-^j\right)^*\left(\tilde{a}_-^j\right)$ and we have also considered the unitarity of the neutrino survival probability, $\sum_{j=1}^3 P^{3\nu}_{ej} = 1$.
As a result, the effective neutrino magnetic moment in the mass basis can be expressed as~\cite{Grimus:2002vb}
\begin{equation}
\label{eq:nmm_sun}
(\mu^{M}_{\rm{sol}})^{2} = |\mathbf{\Lambda}|^{2} -
  c^{2}_{13}|\Lambda_{2}|^{2} + (c^{2}_{13}-1)|\Lambda_{3}|^{2} +
  c^{2}_{13}P^{2\nu}_{e1}(|\Lambda_{2}|^{2}-|\Lambda_{1}|^{2}),
\end{equation}
where $P^{2\nu}_{e1}$ is the effective two-neutrino oscillation probability for solar neutrinos and 
$c_{13}^2 = \cos^2\theta_{13}$. 
Note that this expression does not depend on any of the phases of the TMM matrix, as has already been noticed before~\cite{Grimus:2002vb}\footnote{This expression is also equivalent to the equations presented in Refs.~\cite{Borexino:2017fbd, Khan:2017djo}.}.

\begin{figure}[!t]
\centering
\includegraphics[width=0.49\textwidth]{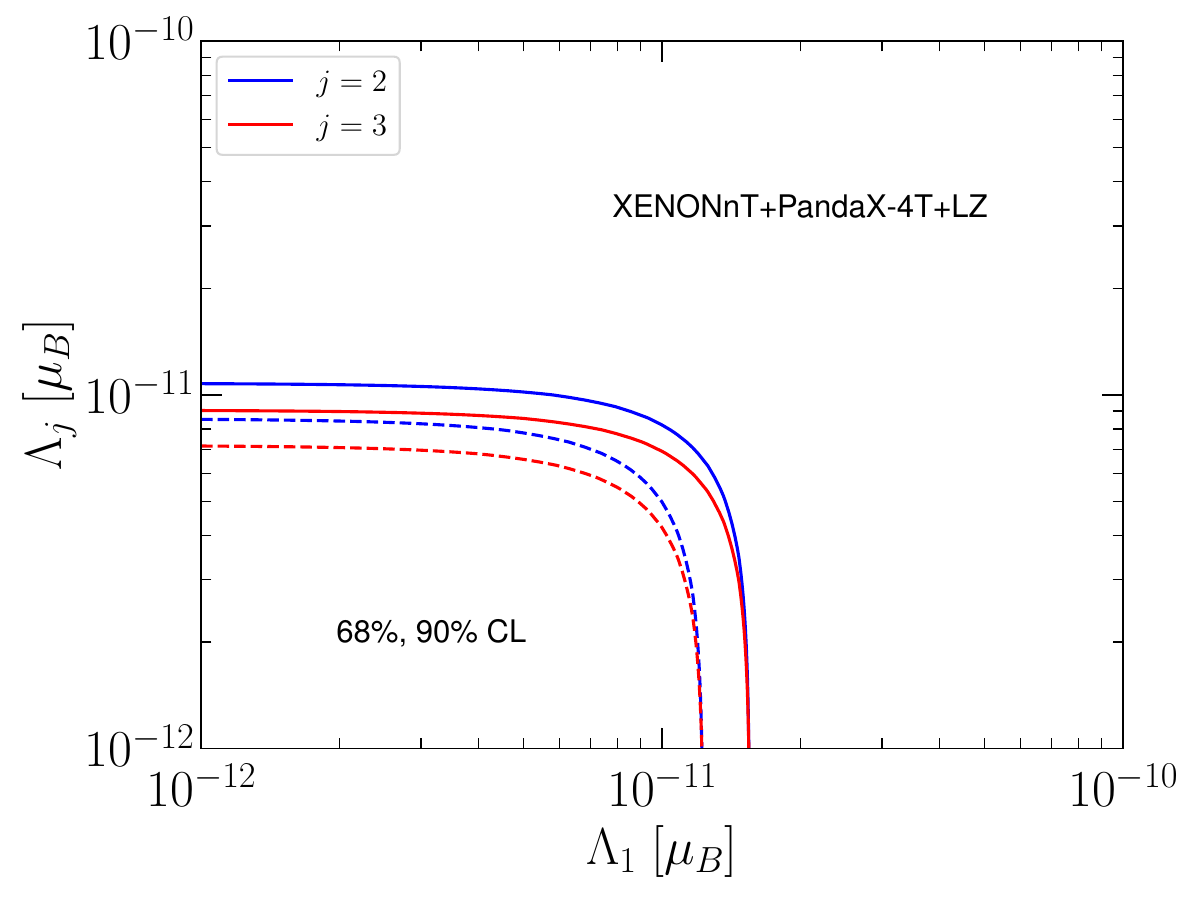}
\includegraphics[width=0.49\textwidth]{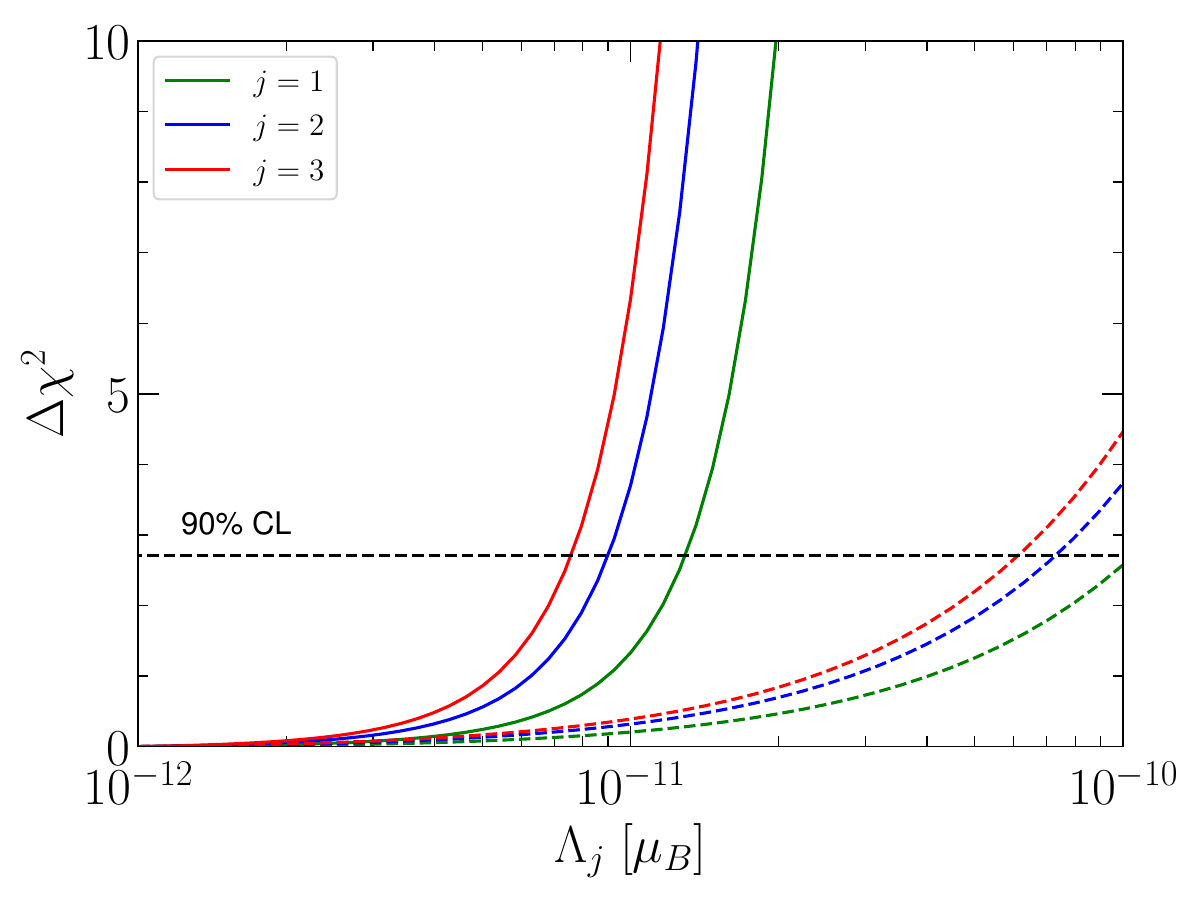}
\caption{Left: The 1$\sigma$ (dashed line)and 90\% (solid line) contours in the $\Lambda_1-\Lambda_j$ plane obtained from DMDD data. Right: The corresponding $\Delta\chi^2$ profiles (solid) for the individual $\Lambda_j$. Also shown for comparison the bounds from Borexino (dashed).}
\label{fig:dmdd}
\end{figure}

In this section, we also examine the bounds derived from the analysis of data from DMDD experiments, in particular XENONnT~\cite{XENON:2022ltv}, LUX-ZEPLIN (LZ)~\cite{LZ:2022lsv} and PandaX-4T~\cite{PandaX:2022ood}. 
Following the approach in~\cite{Giunti:2023yha},
we perform a direct fit to the fundamental parameters rather than to the effective magnetic moment. The result of the combined analysis of all three experiments is shown in Fig.~\ref{fig:dmdd}.
The left panel displays the 1$\sigma$ and 90\% CL allowed regions (for two degrees of freedom) in the $\Lambda_1-\Lambda_2$ (blue) and  $\Lambda_1-\Lambda_3$ (red) planes.
As previously noted, the effective magnetic moment for solar neutrinos is independent of the CP-violating phases, and thus no cancellations can happen among the parameters. 
Overall, the bounds obtained here are stronger than those from the short-baseline experiments GEMMA or LSND, discussed earlier, and they also improve upon  previous solar neutrino limits  from Super-Kamiokande and Borexino .
The right panel shows the $\Delta\chi^2$ profiles for the individual $\Lambda_i$, along with a comparison to Borexino-based bounds~\cite{Coloma:2022umy} (dashed lines). 
These results highlight that the lower energy thresholds of DMDD experiments allow them to set stronger constraints than dedicated solar neutrino experiments.
The bounds from DMDD experiments at 90\% CL read
\begin{eqnarray}
    \Lambda_1 &<& 1.3\times 10^{-11}~\mu_{\text{B}}\,, \\    
    \Lambda_2 &<& 0.9\times 10^{-11}~\mu_{\text{B}}\,, \\    
    \Lambda_3 &<& 0.8\times 10^{-11}~\mu_{\text{B}}\,,    
\end{eqnarray}
and update those reported in Ref.~\cite{AristizabalSierra:2021fuc}
which were based on earlier data from XENON-1T~\cite{XENON:2020rca}. In the next section we will use these updated constraints to infer the corresponding values of the effective magnetic moments relevant for other experimental setups.

\section{Implications for reactor and accelerator experiments}
\label{sec:implications}

\begin{figure}[!t]
\centering
\includegraphics[width=0.49\textwidth]{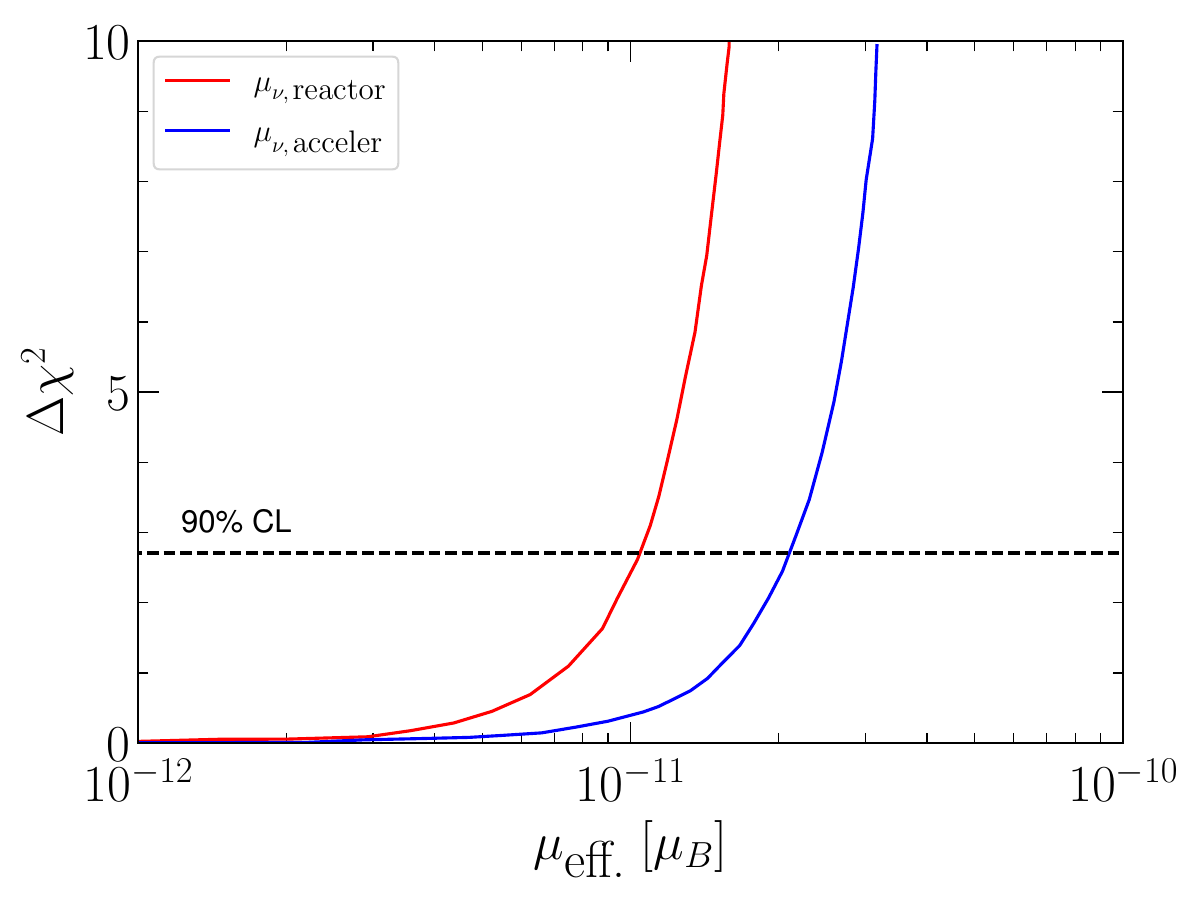}
\caption{Constraints on the effective neutrino magnetic moments in reactor and accelerator experiments, derived from the bounds on the fundamental parameters $\Lambda_i$ obtained in DMDD experiments.}
\label{fig:bound_translate}
\end{figure}

We now turn to the interpretation of our results. We use the stringent bounds on the fundamental parameters $\Lambda_i$, derived from DMDD experiments, to infer the corresponding limits on the effective magnetic moments relevant for SBL reactor and accelerator experiments. These translated constraints are shown in Fig.~\ref{fig:bound_translate}, and serve to illustrate how bounds on the underlying parameters map onto observables in different experimental contexts.
At 90\% confidence level, the bounds read
\begin{eqnarray}
    {\mu_{\nu, \, \text{reactor}}} &<& 1.0\times10^{-11}\mu_{\text{B}} \,, \\
    {\mu_{\nu, \, \text{acceler}}} &<& 2.1\times10^{-11}\mu_{\text{B}} \,.   
\end{eqnarray}
This implies that current and upcoming searches at reactor and accelerator facilities must improve their sensitivity to compete with those based on solar neutrino data. 
In Tab.~\ref{tab:bounds} we summarize several bounds from reactor and accelerator experiments reported in the literature. As can be seen, some reactor experiments are getting close to the translated bound from DMDD experiments, whereas accelerator experiments would require an improvement of about one order of magnitude to become competitive.
The bounds from reactor and accelerator experiments need to be compared to the numbers above and not with the effective magnetic moment obtained from solar neutrinos, which reads $\mu_\text{sol} < 7.5 \times 10^{-12}~\mu_{\text{B}}$~\cite{Giunti:2023yha}.
Nevertheless, it should be noted that, even if future experiments reach similar sensitivities, combined analyses involving  reactor, accelerator and solar neutrino experiments will remain essential. Such a joint approach is crucial to resolve the parameter degeneracies inherent in the definition of $\mu_\nu$ and to eliminate degeneracies in the $\Lambda_i$ parameter space.

\begin{table}[t]
    \centering
    \begin{tabular}{|c|c|c|c|}
\hline
         Channel & Experiment & Bound & ~~Reference~~ \\
         \hline
         ~~Reactor E$\nu$ES~~ & ~~Krasnoyarsk~~ & ~~$\mu_{\nu_e} < 2.4\times 10^{-10}$ (90\% CL)~~ & \cite{Vidyakin:1992nf}\\
          & Rovno & $\mu_{\nu_e} < 1.9\times 10^{-10}$ (95\% CL) & \cite{Derbin:1993wy}\\
          & MUNU & $\mu_{\nu_e} < 9.0\times 10^{-11}$ (90\% CL) & \cite{MUNU:2005xnz}\\
          & TEXONO & $\mu_{\nu_e} < 7.4\times 10^{-11}$ (90\% CL) & \cite{TEXONO:2006xds}\\
          & GEMMA & $\mu_{\nu_e} < 2.9\times 10^{-11}$ (90\% CL) & \cite{Beda:2012zz}\\
          & CONUS & $\mu_{\nu_e} < 7.5\times 10^{-11}$ (90\% CL) & \cite{CONUS:2022qbb}\\
\hline         
         ~~Reactor CE$\nu$NS + E$\nu$ES~~ & ~~Dresden-II~~ & ~~$\mu_{\nu_e} < 2.1\times 10^{-10}$ (90\% CL)~~ & \cite{AtzoriCorona:2022qrf}\\
          & ~~TEXONO~~ & ~~$\mu_{\nu_e} < 2.4\times 10^{-10}$ (90\% CL)~~ & \cite{AtzoriCorona:2025ygn}\\
          & ~~CONUS+~~ & ~~$\mu_{\nu_e} < 2.1\times 10^{-10}$ (90\% CL)~~ & \cite{DeRomeri:2025csu}\\
\hline         
         ~~Accelerator E$\nu$ES~~ & ~~LAMPF~~ & ~~$\mu_{\nu, \text{acceler}} < 6.1\times 10^{-10}$ (90\% CL)~~ & \cite{Allen:1992qe}\\
          & ~~LSND~~ & ~~$\mu_{\nu, \text{acceler}} < 5.7\times 10^{-10}$ (90\% CL)~~ & \cite{LSND:2001akn}\\
          \hline
         ~~Accelerator CE$\nu$NS + E$\nu$ES~~ & ~~COHERENT~~ & ~~$\mu_{\nu, \text{acceler}} < 1.8\times 10^{-9}$ (90\% CL)~~ & \cite{AtzoriCorona:2022qrf}\\
\hline
    \end{tabular}
    \caption{\label{tab:bounds} Bounds on the effective magnetic moment from selected reactor and accelerator experiments. Note that, in the case of COHERENT, a recast for $\mu_{\nu, \text{acceler}}$ is not possible and we simply report the strongest bound from Ref.~\cite{AtzoriCorona:2022qrf} (the one on $\mu_{\nu_\mu}$). The real bound would be slightly stronger.}
\end{table}

\section{Conclusions}
\label{sec:conc}

In this work, we have aimed to clarify persistent misconceptions within the community regarding the interpretation of neutrino magnetic moment bounds. In particular, we have demonstrated how the effective magnetic moment is experiment-dependent, differing in its form for reactor, accelerator, and solar neutrino setups.
We have updated the constraints on the fundamental transition magnetic moments of Majorana neutrinos using the latest results from dark matter direct detection experiments. Thanks to their low energy thresholds, these experiments,
which receive an  irreducible background contribution from solar neutrinos, achieve stronger bounds than dedicated solar neutrino detectors such as Borexino.
These updated bounds on the fundamental parameters were then used to derive the corresponding constraints on the effective magnetic moments relevant for reactor and accelerator neutrino experiments. As shown in Section~\ref{sec:implications}, the effective magnetic moments inferred from DMDD limits differ significantly from the direct bound on $\mu_\text{sol}$. Therefore, it is this translated value—not the solar neutrino limit itself—that should be used as a benchmark when evaluating or comparing reactor and accelerator constraints or experimental sensitivities, contrary to common practice in the literature.
Finally, note that, while our study has focused on transition magnetic moments for Majorana neutrinos, the same analysis can be straightforwardly extended to the Dirac case, with appropriate modifications to account for the different structure of their magnetic moment matrices~\cite{AristizabalSierra:2021fuc}. This highlights the general relevance of our results and the need for careful interpretation of experimental bounds across different neutrino frameworks.

\section*{Acknowledgments}
We would like to thank Dimitris Papoulias and Gonzalo Sánchez García for helpful discussions.
Work supported by the Spanish grants PID2023-147306NB-I00 and CEX2023-001292-S (MCIU/AEI/ 10.13039/501100011033) and CIPROM/2021/054 from Generalitat Valenciana.

\bibliographystyle{utphys}
\bibliography{bibliography}  

\end{document}